# Plasmonic Nanoslit Array Enhanced Metal-Semiconductor-Metal Optical Detectors

Sukru Burc Eryilmaz, Onur Tidin, and Ali K. Okyay

*Abstract*— Metallic nanoslit arrays integrated on germanium metal-semiconductor-metal photodetectors show many folds of absorption enhancement for transverse-magnetic polarization in the telecommunication C-band. Such high enhancement is attributed to resonant interference of surface plasmon modes at the metal-semiconductor interface. Horizontal surface plasmon modes were reported earlier to inhibit photodetector performance. We computationally show, however, that horizontal modes enhance the efficiency of surface devices despite reducing transmitted light in the far field.

*Index Terms*—Photodetectors, plasmons, silicon germanium

## I. INTRODUCTION

THE SEMICONDUCTOR industry has continuously scaled down electronic devices, significantly improving device-level performance. Unfortunately, the same is not true for the electrical information channels that interconnect the devices on a chip or from chip to chip. The high-frequency nature of photonics makes optical interconnects a natural candidate for high-capacity information transfer medium. There is growing interest in high-speed and low-power optical detectors to realize integrated optoelectronics.

Germanium (Ge) has emerged as one of the strongest candidates for photonic materials that are sensitive at telecommunication standard wavelengths and have seamless compatibility with silicon-based electronics. There is an avalanche of interest in demonstrating efficient Ge-based optical detection closely integrated with silicon and silicon-based waveguide technologies [1]–[4].

Metal-semiconductor-metal (MSM) photodetectors provide increased bandwidth and decreased capacitance compared with *p-i-n* photodetectors [5], [6]. A reduction in the finger spacing allows response times in the order of picoseconds [7]. However, this decreases the sensitivity of the device, limiting the amount of light in the active layer of the detector. The active layer of the photodetector should be thin for high-speed operation [8], which restricts the absorbed optical energy. The finger spacing and active layer thickness should be reduced for high-speed operation while maintaining the efficiency of the device. MSM photodetectors with a finger width of 100 nm and a grating pitch of 200 nm have been demonstrated to provide high bandwidth in devices with a GaAs active layer that is 40 nm thick [9], [10].

Surface plasmons can be utilized to increase absorption in the thin active regions of high-speed optical detectors. Sub-wavelength metallic apertures transmitting an extraordinary amount of light have been a great source of interest. After Ebbesen et al. first demonstrated high transmission through hole-arrays [11], slit arrays have been studied by several authors [12]–[14]. Electromagnetic resonance modes in MSM structures have been investigated, and absorption enhancement in MSM photodetectors with 8-nm-thick single-quantum-well regions have been observed [15]. On the other hand, earlier reports in the literature claim very limited or no advantage at all from the use of metallic gratings for bulk photodetector performance [14], [16]. Our study shows, however, that significant performance improvements can be possible by carefully designing metallic gratings integrated with MSM photodetectors. In this letter, we consider electromagnetic fields in both the near field and the far field, and contributions from surface plasmon oscillation modes to analyze improved detector performance despite the reduced transmitted far-field intensity.

## II. SIMULATION AND DISCUSSION

For a proof-of-concept demonstration in near infrared (NIR) photodetectors, we computationally studied Ge-based MSM photodetectors integrated with gold gratings, as shown in Fig. 1. In our analyses, we computed the field maps of metallic gratings integrated on Ge-based photodetectors and investigated the relative contribution to optical absorption enhancement with respect to the bare device using the two-dimensional finite-difference time-domain method (Lumerical Solutions Inc., Canada). In these numerical simulations, we calculated the frequency domain response by taking the Fourier transform of time domain representations. This computational approach allowed us to use experimental refractive index data to represent the device materials, including Ge [17] and Au [18] in our structures. The simulation domain boundary conditions along the *x* axis were

Manuscript received July 28, 2011. This work was supported by TUBITAK 108E163, 109E044, EU FP7 PIOS 239444. The authors acknowledge the Turkish Ministry of Industry and Trade seed fund and TUBITAK BIDEP.

The authors are with the Department of Electrical and Electronics Engineering, UNAM-Institute of Materials Science and Nanotechnology, Bilkent University, Ankara 06800, Turkey (e-mail: eryilmaz@stanford.edu; aokyay@stanfordalumni.org).

This is the preprint version of the accepted manuscript with DOI: 10.1109/LPT.2012.2183342. Full citation is as follows: Eryilmaz, S.B.; Tidin, O.; Okyay, A.K., "Plasmonic Nanoslit Array Enhanced Metal–Semiconductor–Metal Optical Detectors," *Photonics Technology Letters, IEEE* , vol.24, no.7, pp.548,550, April1, 2012

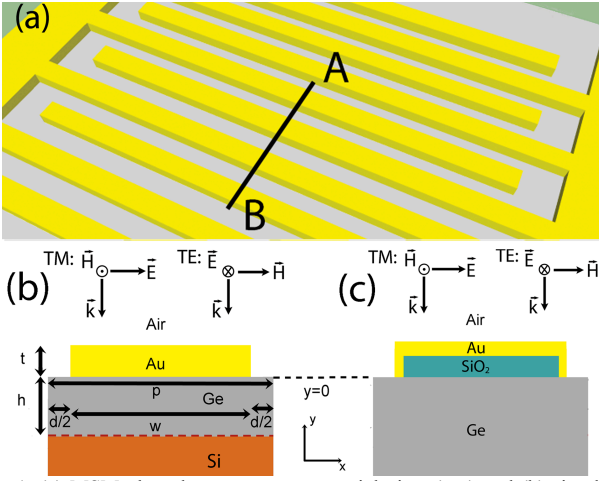

Fig. 1. (a) MSM photodetector structure aerial view (top) and (b) simulated structure for allowed HSP modes with structural parameters. Absorption is calculated for the region between $y = 0$ and $-h$ (active region). (c) Simulated structure for suppressed HSP modes.

set to be periodic, while those along the $y$ axis were set to be perfectly matched.

We first studied the contribution of vertical cavity modes and horizontal surface plasmon (HSP) modes to the transmission and reflection of the incident light. We investigated the far-field and near-field cases separately by calculating the field intensity profiles allowing HSPs (Fig. 1(b)) and suppressing HSPs (Fig. 1(c)), using the metallic grating design with $(d, P, t) = $ (75 nm, 1000 nm, 260 nm), which is the optimum structure for maximum absorption enhancement, as explained below. The structure to suppress HSP modes shown in Fig. 1(c) is similar to the structure used by Crouse et al. for the same purpose [14]. The $SiO_2$ slab in Fig. 1(c) is 150 nm thick and 825 nm wide. The transmission rates for different optical wavelengths were computed for surface normal illumination and transverse-magnetic (TM) and transverse-electric (TE) polarizations by normalizing the total field intensity at 1 μm below the metallic grating to the intensity of the source assuming a non-dispersive and lossless Ge region, in order to decouple absorption in the semiconductor. Figure 2(a) shows the normalized transmission versus the wavelength of the light with and without suppressing HSPs. Surface plasmon oscillations tend to reduce the transmitted far-field intensity except for a narrow range of wavelengths, as observed by Crouse et al. [14]. However, MSM and other high-speed photodetectors are essentially surface devices with thin active layers, and the field intensity in the thin active layer is crucial. *The near-field intensity in the vicinity of the surface of such structures is dramatically different, in favor of HSPs, as shown in Fig. 2(b) and 2(d) (allowing HSPs) and 2(c) and 2(e) (suppressing HSPs).* The standing wave pattern of the field at the surface is due to the interference of surface plasmon oscillations from adjacent gratings and the ends of the metallic slits. Excitation of HSPs at the metal-semiconductor interface significantly increases the field density in the thin active region of the MSM photodetector close to the surface of the device.

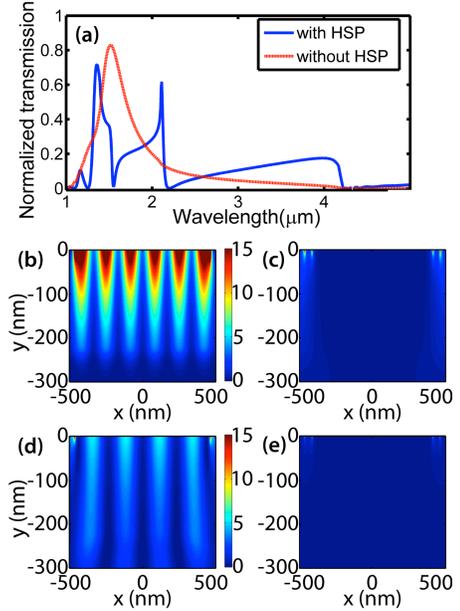

Fig. 2. (a) Far-field transmission (normalized to source power) with and without HSP modes, (b) $E$-field map for $\lambda_i$ = 1550 nm and allowed HSP modes, (c) suppressed HSP modes, (d) $E$-field map for $\lambda_i$ = 2110 nm and allowed HSP modes, and (e) suppressed HSP modes. $E$-field intensity values are clipped at 15 in the color scale. $E$-field intensity is normalized to the $E$-field intensity of the source. All field maps use the same color scale.

We comparatively studied MSM photodetector performance with metallic slits and the case of the starting photodetector with no metallic gratings (dubbed "bare" device) by parametrically optimizing the metallic slits for maximum absorption enhancement ($A$) in the active region, assuming that the internal quantum efficiency of the photodetector remains unchanged. The overall absorptivity is calculated in the active Ge layer by

$$\omega \times \text{Im}(\varepsilon) \cdot \oint_V |E|^2 dV \quad (1)$$

where $E$ is the electric field, $V$ is the volume where the absorption is to be calculated, and $\varepsilon$ is the dielectric constant of the material filling the volume. Here, the optical coefficients of strained Ge [17] grown on silicon are used for real and imaginary refractive indices of the Ge material. The devices are illuminated with normally incident planewaves with TM-polarized illumination, i.e., $A_{TM}$, and TE-polarized illumination, i.e., $A_{TE}$. We calculated the absorption in the active Ge layer between $y = 0$ and $-h$. Absorption enhancement for the metallic structures with structural parameters $(d, P, t)$ in Fig. 1(b) is computed as the ratio of overall absorption of the Ge layer with metallic structure, to the absorption of the same Ge layer with no metallic structure (bare device). An 8-fold (for $h = 300$ nm) maximum absorption enhancement at the illumination wavelength of $\lambda_i =$ 1550 nm is obtained for $(d, P, t) = $ (75 nm, 1000 nm, 260 nm). Figure 3(a) shows the absorption enhancement, for the same value of $h$, with respect to $P$ and $d$ when $t = 260$ nm. The surface plasmon wavelength at 1550 nm illumination is calculated to be $\lambda_s = 327$ nm, which is almost the spacing between pitch values corresponding to enhancement peaks in Fig. 3(a). For $P = 1000$ nm and $t = 260$ nm, the enhancement changes rapidly with $d$ between 25 and 200 nm, as shown in Fig. 3(b). The strong dependence of enhancement on metal thickness, as well as a number of enhancement peaks in Fig. 3(b), can be attributed to the influence of cavity modes. Increasing the width of the slits beyond 100 nm reduces the total absorbed energy in the semiconductor despite the reduced metal coverage. The silicon layer is omitted in these computations, because silicon is transparent at 1550 nm and a significant portion of the light is absorbed before it reaches silicon. For our optimum design, we repeat the computation including a semi-infinite silicon layer located below $y = -h$ (silicon substrate), while keeping the simulation boundaries the same. The total absorption of the bare structure decreases by 9%, while that of the structure with the optimized grating decreases by 0.4%, implying that Fresnel reflections occurring at the germanium-silicon interface result in a negligible change in the absorption calculations.

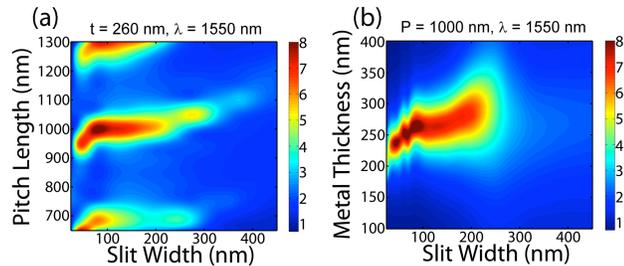

Fig. 3. (a) Absorption enhancement with respect to pitch length, $P$, and slit width, $d$. (b) Absorption enhancement with respect to metal thickness, $t$, and slit width, $d$. Enhancement values are clipped at 8 in the color scale.

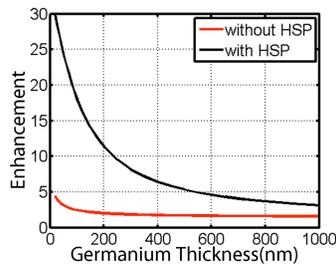

Fig. 4. Absorption enhancement for the optimum structure with respect to active Ge layer thickness with and without HSP modes

Absorption enhancement is greater for photodetectors with active areas thinner than 300 nm. Figure 4 shows the computed absorption enhancement factors for different active area thicknesses comparatively for cases allowing and suppressing HSPs. In these simulations, $h$ varies from 10 nm to 1 μm. For active layers thinner than 100 nm, the enhancement factor exceeds 20× with the presence of HSPs, due to strong fields at the surface. HSP modes enhance the absorption for photodetectors with active layer thicknesses even greater than 1 μm, complementing the claims in related reports on the role of HSP oscillations [14].

Absorption in the semiconductor is significantly inhibited for TE-polarized illumination, because such illumination cannot excite surface plasmons and is largely reflected from the metallic structure. For TE-polarized illumination, absorption in the semiconductor further decreases with increased metal thickness and finger width. For $w = 900$ nm and $P = 1000$ nm, we observe that absorption is below 2% of the bare Ge case. Therefore, the overall enhancement factor for unpolarized illumination $(A_{TM} + A_{TE})/2$ is around 4×, assuming that any contribution to the absorption enhancement from TE-polarized light is negligible.

### III. CONCLUSION

We analyzed and utilized surface plasmon modes for the design of Ge MSM photodetectors operating in telecommunication bands. An absorption enhancement factor up to 8× was obtained. The results are interpreted as the combined effect of horizontal surface plasmon modes and vertical cavity modes.

REFERENCES

[1] M. Oehme, J. Werner, E. Kasper, M.Jutzi, and M. Berroth, "High bandwidth Ge p-i-n photodetector integrated on Si," *Appl. Phys. Lett.* Vol. 89, no. 7, pp. 071117-071117-3, Feb. 2006.
[2] Y. Kang et al., "Monolithic germanium/silicon avalanche photodiodes with 340 GHz gain-bandwidth product," *Nature Photon.*, vol. 3, pp. 59-63, Jan. 2009.
[3] J. Michel, J. Liu, and L. C. Kimerling, "High-performance Ge-on-Si photodetectors," *Nature Photon.*, vol. 4, pp. 527-534, Aug. 2010.
[4] S. Assefa, F. Xia, and Y. A. Vlasov, "Reinventing germanium avalanche photodetector for nanophotonic on-chip optical interconnects," *Nature*, vol. 464, pp. 80-85, Mar. 2010.
[5] J. B. D. Soole and H. Schumacher, "InGaAs metal-semiconductor-metal photodetectors for long wavelength optical communications," *IEEE J. Quantum Electron.*, vol. 27, no. 3, pp. 737-752, Mar. 1991.
[6] M. Ito and O. Wada, "Low dark current GaAs metal-semiconductor-metal (MSM) photodiodes using WSixcontacts," *IEEE J. Quantum Electron.*, vol. 22, no. 7, pp. 1073-1077, Jul. 1986.
[7] S. Y. Chou, and M. Y. Liu, "Nanoscale tera-hertz metal-semiconductor-metal photodetectors," *IEEE J. Quantum Electron.*, vol. 28, no. 10, pp. 2358-2368, Oct. 1992.
[8] D. A. B. Miller, "Optical interconnects to electronic chips," *Appl. Opt.*, vol. 49, no. 25, pp. 59-70, Sep. 2010.
[9] S. Collin, F. Pardo, and J.-L. Pelouard, "Resonant-cavity-enhanced subwavelength metal-semiconductor-metal photodetector," *Appl. Phys. Lett.*, vol. 83, no. 8, pp. 1521-1523, Aug. 2003.
[10] S. Collin, F. Pardo, R. Teissier, and J.-L. Pelouard, "Efficient light absorption in metal-semiconductor-metal nanostructures," *Appl. Phys. Lett.*, vol. 85, no. 2, pp. 194-196, Jul. 2004.
[11] T. W. Ebbesen, H. J. Lezec, H. F. Ghaemi, T. Thio, and P. A. Wolff, "Extraordinary optical transmission through sub-wavelength hole arrays," *Nature*, vol. 391, pp. 667-669, Feb. 1998.
[12] S. Collin, F. Pardo, R. Teissier, and J. L. Pelouard, "Horizontal and vertical surface resonances in transmission metallic gratings," *J. Opt. A: Pure Appl. Opt.*, vol. 4, no. 5, pp. 154-160, Sep. 2002.
[13] J. A. Porto, F. J. García-Vidal, and J. B. Pendry, "Transmission resonances on metallic gratings with very narrow slits," *Phys. Rev. Lett.*, vol. 83, no. 14, pp. 2845-2848, Oct. 1999.
[14] D. Crouse and P. Keshavareddy, "Role of optical and surface plasmon modes in enhanced transmission and applications," *Opt. Express*, vol. 13, no. 20, pp. 7760-7771, Oct. 2005.
[15] J. Hetterich *et al.*, "Optimized design of plasmonic MSM photodetector," *IEEE J. Quantum Electron.* vol. 43, no. 10, pp. 855-859, Oct. 2007.
[16] D. Crouse and P. Keshavareddy, "A method for designing electromagnetic resonance enhanced silicon-on-insulator metal-semiconductor-metal photodetectors," *J. Opt. A: Pure Appl. Opt.*, vol. 8, no. 2, pp. 175-181, Feb. 2006.
[17] A. K. Okyay, "Si-Ge photodetection technologies for integrated optoelectronics," Ph.D. dissertation, Dept. Elect. Eng., Stanford Univ., Stanford, CA, 2007.
[18] E. D. Palik, *Handbook of Optical Constants of Solids*. New York: Academic, 1985, pp. 294.